\documentclass[pre,showpacs,oneocolumn,amsfonts,amssymb,amsmath]{revtex4}
\usepackage{amsmath}
\usepackage{graphicx}
\usepackage{subfigure}
\usepackage{color}

\def \be {\begin{equation}}
\def \ee {\end{equation}}

\def \rv {\ensuremath{\mathbf{r}}}
\def \dr {\ensuremath{\text{d}\mathbf{r}}}

\def \kv {\ensuremath{\mathbf{k}}}

\begin{document}

\title{Pearson Walk with Shrinking Steps in Two Dimensions}

\author{C. A. Serino}
\affiliation{Center for Polymer Studies and Department of Physics, Boston
  University, Boston, MA, USA~ 02215}
\email{cserino@physics.bu.edu}
\author{S. Redner} 
\affiliation{Center for Polymer Studies and Department of Physics, Boston
  University, Boston, MA, USA~ 02215}
\email{redner@bu.edu}


\begin{abstract}
  We study the shrinking Pearson random walk in two dimensions and greater,
  in which the direction of the $N^{\rm th}$ step is random and its length equals
  $\lambda^{N-1}$, with $\lambda<1$.  As $\lambda$ increases past a critical
  value $\lambda_c$, the endpoint distribution in two dimensions,
  $P(\mathbf{r})$, changes from having a global maximum away from the origin
  to being peaked at the origin.  The probability distribution for a single
  coordinate, $P(x)$, undergoes a similar transition, but exhibits multiple
  maxima on a fine length scale for $\lambda$ close to $\lambda_c$.  We
  numerically determine $P(\mathbf{r})$ and $P(x)$ by applying a known
  algorithm that accurately inverts the exact Bessel function product form of
  the Fourier transform for the probability distributions.
\end{abstract}
\maketitle

\section{Introduction}
\label{sec.intro}

\begin{figure}[ht]
\begin{center}
\includegraphics[width=0.35\textwidth]{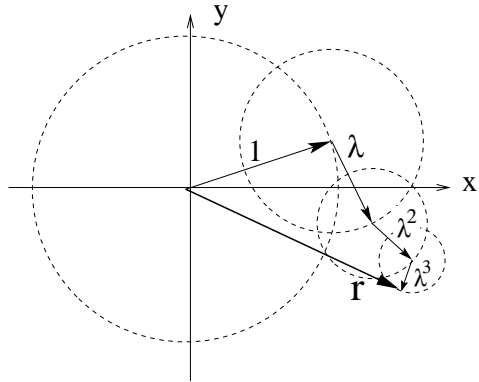}
\caption{Illustration of the first four steps of a shrinking Pearson walk
  in two dimensions, leading to a displacement $\mathbf{r}$.}
\label{fig:model}
\end{center}
\end{figure}

In this work, we investigate the probability distribution of the {\em
  shrinking Pearson random walk\/} in two and greater dimensions, in which
the length of the $N^{\rm th}$ step equals $\lambda^{N-1}$, with $\lambda<1$.
If a walk is at $\mathbf{r}_N$ after the $N^{\rm th}$ step, then
$\mathbf{r}_{N+1}$ is uniformly distributed on the surface of a sphere of
radius $\lambda^N$ centered about $\mathbf{r}_N$ (Fig.~\ref{fig:model}).  We
assume that the walk begins at the origin, and the length of the first step
is $\lambda^0=1$.  The random direction for each step corresponds to the
classic Pearson walk~\cite{pearson,W94}, whose solution is well known when
the length of each step is fixed.  In this case, the central limit theorem
guarantees that the asymptotic probability distribution of endpoints
approaches a Gaussian function.

In one dimension, the random walk with exponentially shrinking step lengths
exhibits a variety of beautiful properties~\cite{W,E}.  For
$\lambda<\frac{1}{2}$, the support of the endpoint distribution after $N$
steps, $P_N(x)$, is a Cantor set, while for $\lambda>\frac{1}{2}$ the support
is the connected interval $[-\frac{1}{1-\lambda},\frac{1}{1-\lambda}]$.  More
interestingly, for $\frac{1}{2}<\lambda<1$ and for $N\to\infty$, $P_N(x)$ is
continuous for almost all values of $\lambda$, but is fractal on a
complementary and infinite discrete set of $\lambda$ values~\cite{W,E,G,Kac}.
A particularly striking special case is $\lambda=g \equiv
\frac{1}{2}(\sqrt{5}-1)=0.618\ldots$ (the inverse of the golden ratio), where
$P(x)$ is artistically self-similar on all length scales~\cite{KR,PSS}.

Shrinking random walks in greater than one dimension are much less studied.
The probability distribution of short Pearson walks with a step size that
decays as a power law in the number of steps was treated by Barkai and
Silbey~\cite{BS99}, while the probability distribution of short Pearson walks
with arbitrary unequal step sizes was considered by Weiss and
Kiefer~\cite{WK83}.  More recently, Rador~\cite{R06} studied the moments and
various correlations of the probability distribution, and also developed a
$1/d$ expansion method, where $d$ is the spatial dimension, for Pearson walks
with shrinking steps.

A physical motivation for this model comes from granular media.  If a
granular gas is excited and then allowed to relax to a static state, the
motion of a labeled particle is equivalent to a random walk whose steps
lengths decrease because of the loss of energy by repeated inelastic
collisions.  A related example is an inelastic ball that is bouncing on a
vibrating platform~\cite{MK07}, where the velocity of the ball after each
bounce essentially experiences a random walk with shrinking steps if the
vibration is sufficiently weak.  Our interest was prompted by
M. Bazant~\cite{bazant}, who apparently introduced the shrinking Pearson walk
in an MIT graduate mathematics course on random walks.

While the distribution of radial displacements, $P({r})$, no longer exhibits
self-similar properties, numerical simulations indicated that $P({r})$
qualitatively changes shape as a function of $\lambda$.  For $\lambda\ll 1$,
the support of $P({r})$ is confined to
$1-\tfrac{\lambda}{1-\lambda}<r<1+\tfrac{\lambda}{1-\lambda}$, and the
distribution is peaked near $r=1$.  As $\lambda$ increases beyond
$\tfrac{1}{2}$, the probability of being near the origin increases and
$P({r})$ eventually exhibits a maximum at the origin when $\lambda$ exceeds a
critical value, $\lambda_c(r)$.  For two spatial dimensions, we estimate
$\lambda_c(r)$ to be $0.5753882\pm 0.0000003$.

The distribution of a single coordinate, $P(x)$, undergoes a similar shape
transition, but at a slight different critical value, $\lambda_c(x)$, that we
estimate to be $0.558458\pm 0.000003$.  More surprisingly, $P(x)$
exhibits up to seven local minima and maxima when $\lambda\approx
\lambda_c(x)$.  The secondary extrema occur on a very fine scale that can be
resolved only by a high-accuracy numerical method, due to Van Deun and
Cools~\cite{vdc}, to invert the Fourier transform of the probability
distribution.

In the next section, we present some elementary properties of the shrinking
Pearson random walk and show how to obtain the exact Fourier transform for
the radial and single-coordinate probability distributions.  In Sec.~III, we
apply the Van Deun and Cools algorithm to numerically invert the Fourier
transform with high accuracy.  From this inversion, we outline the behaviors
of the radial and single-coordinate probability distributions as a function
of $\lambda$ in Sec.~IV.  We briefly discuss the shrinking Pearson walk in
spatial dimensions $d>2$ in Sec.~V and conclude in Sec.~VI.

\section{Basic Properties}
\label{sec.mgf}

When the length of the $N^{\rm th}$ step decreases exponentially with $N$,
the shrinking Pearson walk eventually comes to a stop at a finite distance
from its starting point.  Since the direction of successive steps are
uncorrelated, the mean-square displacement after the $N^{\rm th}$ step,
$\left<r^2\right>_{N}$, is given by:
\begin{eqnarray}
  \left<r^{2}\right>_{N} &=&
  \left<[\mathbf{r}_1+\mathbf{r}_2+\mathbf{r}_3+\ldots+\mathbf{r}_{N}]^{2}\right>
  = [r_1^{2}+r_2^{2}+r_3^{2}+\ldots+r_{N}^{2}]\nonumber \\
  &=& \left[1 + \left(\lambda\right)^{2}+(\lambda^2)^{2}+\ldots+(\lambda^{N-1})^{2}\right]
  = \frac{1-\lambda^{2N}}{(1-\lambda^{2})}\longrightarrow  \frac{1}{(1-\lambda^{2})}\quad N\to\infty.
\end{eqnarray}
In the second line, we use the fact that the directions of different steps
are uncorrelated so that the average value of all cross terms in the expansion of
$[\mathbf{r}_1+\mathbf{r}_2+\mathbf{r}_3+\ldots+\mathbf{r}_{N}]^{2}$ vanish.
We thus obtain the obvious result that $\left<r^{2}\right>_{N}$ grows
monotonically with $\lambda$ and diverges as $\lambda\to 1$, corresponding to
the infinite-time limit of the classic Pearson random walk.

Our interest is in the probability distributions of the radial coordinate and
a single Cartesian component after $N$ steps, $P_N({r})$ and $P_N(x)$,
respectively, as well as their $N\to\infty$ limiting forms, $P({r})$ and
$P(x)$.  These two distributions undergo a transition from being peaked away
from the origin for small $\lambda$, to being peaked at the origin for
$\lambda$ greater than a critical value.  A transition from a unimodal to
bimodal probability distribution can be constructed, for example, from
Brownian motion in media with non-linear shear profiles~\cite{NL}.  Here the
competition between the flow and diffusion drive the transition.  In the
prsent example, the transition is purely statistical in origin.

\begin{figure}[ht]
\begin{center}
\includegraphics[width=0.95\textwidth]{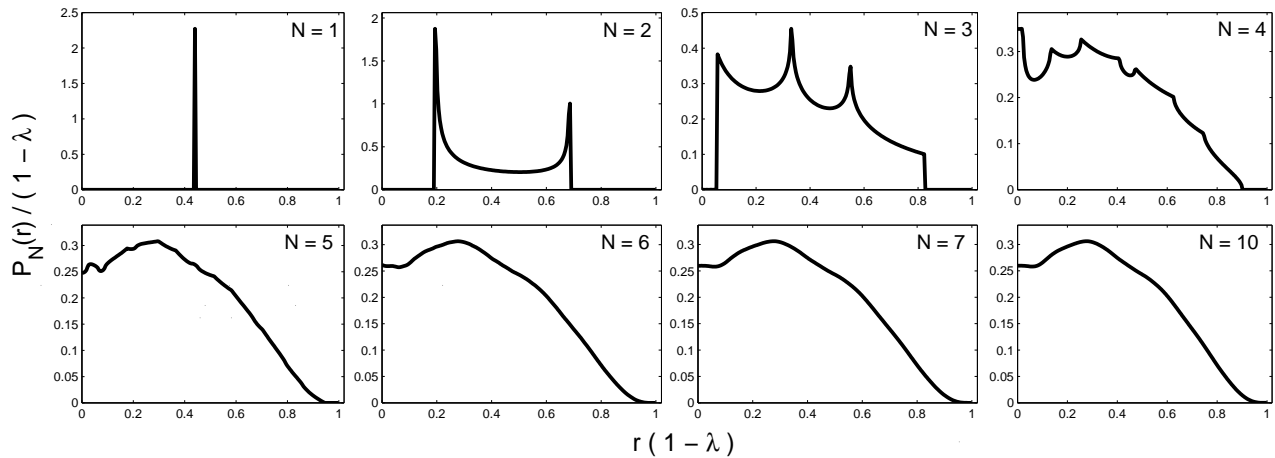}
\caption{The radial distribution $P_N(r)$ for the representative case
  $\lambda=0.56$, with $N=1$, 2, 3, 4, 5, 6, 7, and 10 steps (upper left to
  lower right). }
\label{fig:r}
\end{center}
\end{figure}

Figure~\ref{fig:r} shows the radial distribution for
$\lambda\approx\lambda_c$ after a small number of steps to provide a sense
for the convergence rate to the asymptotic form.  For convenience in putting
many panels on the same scale, we typically plot the distribution $P(r)r_{\rm
  max}$ versus $r/r_{\rm max}$, where $r_{\rm max}=(1-\lambda)^{-1}$ is the
maximal displacement of the infinite walk.  Already by $N=7$ steps, the
probability distribution is visually indistinguishable from its asymptotic
form.  While $P(r)$ varies smoothly as a function of $\lambda$, the position
of the global maximum changes {\it discontinuously} from being peaked at
$r>0$ to being peaked at $r=0$ as $\lambda$ increase beyond a critical value
$\lambda_c(r)$.  The single-coordinate distribution $P(x)$ exhibits a
transition from multimodality to unimodality that somewhat resembles the
transition for $P(r)$, but is more complex in its microscopic details.

Conventionally, the distribution of the displacment factorizes into a product
of single-coordinate distribution, from which the radial distribution follows
easily.  However, in contrast to the classic Pearson walk in which the length
of each step is the same, the probability distribution for the shrinking
Pearson walk no longer factorizes as $P(\mathbf{r})=P(x)P(y)$.  The
differences between the radial and single-coordinate distributions arise
because there is a non-trivial correlation between steps in orthogonal
directions.  If the endpoint of the walk is close to its maximum possible
value in, say, the $x$-direction, then the displacement in the $y$-direction
is necessarily small, and {\it vice versa}.

It is worth emphasizing that it is not practical to accurately determine the
probability distribution of the Pearson random walk with shrinking steps by
straightforward simulations.  As we shall see, the nature of the transition
in $P(x)$ is delicate.  It would require a prohibitively large number of
walks, or a prohibitively fine spatial grid in an exact enumeration method,
to obtain sufficient accuracy to resolve these subtle features.  For this
reason, we employ an alternative approach that is based on calculating the
Fourier transform of the probability distribution --- which can be done
exactly by elementary methods --- and then inverting this transform by the
highly accurate Van Deun and Cools~\cite{vdc} algorithm.

\section{Fourier Transform Solution of the Probability Distribution}

\subsection{Single-Coordinate Distribution}
\label{sec:scd}

We first study the distribution of the (horizontal) $x$ coordinate.  To
obtain the distribution of $x$ after $N$ steps, $P_N(x)$, we start with the
Chapman-Kolmogorov equation~\cite{W94} that relates $P_N(x)$ to $P_{N-1}(x)$,
\begin{equation}
\label{eq.eom}
P_N(x)=\int dx'\:P_{N-1}(x')\, q_N(x-x'),
\end{equation} 
where $q_N(w)$ is the probability of making a displacement whose horizontal
component equals $w$ at the $N^{\rm th}$ step.  Equation~\eqref{eq.eom}
states that to reach a point whose horizontal component equals $x$ after $N$
steps, the walk must first reach a point with horizontal component $x'$ in
$N-1$ steps and then hop from $x'$ to $x$ at the $N^{\rm th}$ step.

We now introduce the Fourier transforms
\begin{eqnarray*}
  P_N(k)= \int dx\:P_N(x)\:e^{ikx}~, \qquad\mathrm{and} \qquad
  q_N(k) = \int dx\:q_N(x)\:e^{ikx}~, 
\end{eqnarray*}
to recast the convolution in Eq.~\eqref{eq.eom} as the product
$P_N(k)=P_{N-1}(k)\,q_N(k)$.  This equation has the formal solution
\begin{equation}
\label{PNk}
  P_N(k)=P_0(k)\prod_{n=0}^{N}q_n(k) = \prod_{n=0}^{N}q_n(k)\,.
\end{equation}
The latter equality applies for a walk that begins at the origin, so that
$P_0(k)=1$.  Now $q_n(x)$ may be obtained by transforming from the uniform
distribution of angles to the distribution of the horizontal coordinate in a
single step by using the relation
\begin{equation}
\label{tform}
q_n(x)\, dx = q_n(\theta)\, d\theta = \frac{d\theta}{2\pi}~,
\end{equation}
together with $x=\lambda^{n-1}\cos\theta$, to give
\begin{equation}
\label{qN}
q_n(x) = \frac{1}{\pi}\frac{1}{\sqrt{\lambda^{2(n-1)}-x^2}}~.
\end{equation}
Although the distribution of angles is uniform, the single-step distribution
for the $x$-coordinate at the $n^{\rm th}$ step has a ``smile'' appearance,
with maxima at $x=\pm \lambda^{n-1}$ and a minimum at $x=0$.  The probability
distribution of the horizontal coordinate after $N$ steps is a convolution of
these smile functions at different spatial scales.  It is this superposition
that gives $P(x)$ its rich properties for $\lambda \approx \lambda_c$.

Using the transformation between $x$ and $\theta$ in Eq.~\eqref{tform}, the Fourier
transform of the single-step probability is
\begin{eqnarray}
q_n(k) = \int dx\:q_n(x)\:e^{ikx}= 
\frac{1}{2\pi}\int_0^{2\pi}  d\theta\, e^{ik\lambda^{n-1}\cos\theta} = J_0(k\lambda^{n-1})\,,
\end{eqnarray}
where $J_0$ is the Bessel function of the first kind of order zero.  This
result relies on a standard representation of the Bessel function as a
Fourier integral~\cite{AS}.
Thus the Fourier transform of the probability distribution in Eq.~\eqref{PNk}
may be expressed as the finite product of Bessel functions
\begin{equation}
\label{eq.soln}
  P_N(k)=\prod_{n=0}^{N-1} J_0(k\lambda^{n})\,.
\end{equation}
To calculate $P_N(x)$ requires inverting the Fourier transform,
\begin{eqnarray}
\label{PNx}
P_N(x)=\frac{1}{2\pi}\int_{-\infty}^{\infty}dk\: e^{-ikx}P_N(k) 
=\frac{1}{\pi}\int_{0}^{\infty}dk\:\cos kx\prod_{n=0}^{N-1} J_0(k\lambda^{n}),
\end{eqnarray}
where we use the fact that $P_N(k)$ is even in $k$ to obtain the second
equality.

Each of the factors $J_0$ in the product in Eq.~\eqref{PNx} is an oscillatory
function of $k$, and the product itself oscillates more rapidly as the number
of terms $N$ increases.  The evaluation of integrals with such rapidly
oscillating integrands has been the subject of considerable research
\cite{oss}; in particular, integrals of products of Bessel functions appear
in nuclear physics \cite{b3}, quantum field theory \cite{b1}, scattering
theory \cite{b2}, and speech enhancement software \cite{b4}.  Recently, Van
Duen and Cools~\cite{vdc} developed an algorithm that can numerically
calculate integrals of power laws multiplied by a product of Bessel functions
of the first kind quickly and with absolute errors of the order of
$10^{-16}$.  We use their algorithm to compute the probability distribution
$P_N(x)$ with this degree of accuracy.  To implement their approach, we first
write~\cite{AS1}
\begin{equation*}
\cos z = \sqrt{\frac{\pi}{2z}}\left(\frac{1}{\sqrt{z}}\:J_{1/2}(z)-\sqrt{z}\:J_{3/2}(z)\right)
\end{equation*}
to express the right-hand side of Eq.\eqref{PNx} in terms of products of
Bessel functions and a power law only.  With this preliminary, we can
directly apply the Van Duen-Cools algorithm to determine $P_N(x)$ accurately.

\subsection{Radial Distribution}

For the distribution of the radial coordinate $r$, $P_N(r)$, we again start
with the Chapman-Kolmogorov equation~\cite{W94}
\begin{equation}
 P_N(r) = \int d\rv'\:  P_{N-1}(\rv')\mathcal{Q}_N(\rv-\rv'),
\label{eq:eomr}
\end{equation}
where $\mathcal{Q}_N(\mathbf{z})$ is the probability that the walk makes a
vector displacement $\mathbf{z}$ at the $N^{\rm th}$ step, and we use the
angular symmetry of the walk to write $P_N$ as a function of only the
magnitude of the displacement.  Since all angles for the $N^{\text{th}}$ step
are equiprobable,
\begin{equation}
  \mathcal{Q}_N(\rv) = \frac{1}{2\pi r}\:\delta(\lambda^{N-1}-|\rv|),
\end{equation}
where $\delta(x)$ is the Dirac delta function.  Once again, we use the
Fourier transform to reduce the convolution in Eq.~\eqref{eq:eomr} to a
product.  This recursion has the solution
\begin{equation}
P_N(k)=P_0(\kv)\prod_{n=0}^{N-1} J_0(k\lambda^{n})=\prod_{n=0}^{N-1} J_0(k\lambda^{n})\,,
 \label{eq:fsrd}
\end{equation}
with the last equality appropriate for a walk that starts from the origin.
While Eqs.~\eqref{eq.soln} and \eqref{eq:fsrd} are identical, the
corresponding distributions in real space are distinct.  To obtain $P_N(r)$,
we must calculate
\begin{equation}
P_N(\rv) = \frac{1}{(2\pi)^2}\int d\kv\:e^{-i\kv\cdot\rv}P_N(\kv)\,.
\end{equation}
Since $P_N(\kv)$ is a function of the magnitude of $\kv$ only, we can write the
integration in polar coordinates and perform the angular integration to
obtain the spherically symmetric result
\begin{equation}
  P_N(r) =  \frac{1}{2\pi}\int_{0}^{\infty}dk\:k\:J_0(kr)\prod_{n=0}^{N-1}
  J_0(k\lambda^{n})\,.
\label{eq.PNr}
\end{equation} 
In this Bessel product form, we can again apply the Van Duen-Cools
algorithm~\cite{vdc} to invert this Fourier transform numerically.

\section{THE PROBABILITY DISTRIBUTIONS}

\begin{figure}[ht]
\begin{center}
\includegraphics[width=0.8\textwidth]{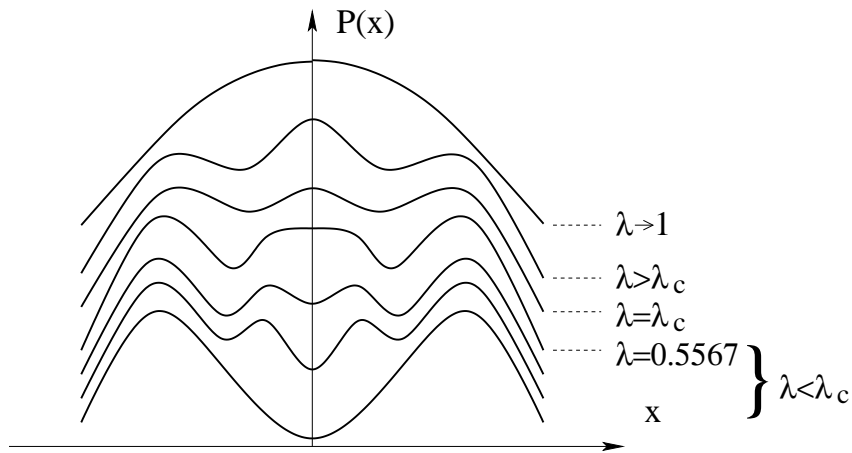}
\caption{ Schematic and not to scale form of $P(x)$ for increasing $\lambda$
  near $\lambda_c$ (bottom to top).  For $\lambda\approx 0.5567<\lambda_c$
  the curvature at the origin becomes positive, while at $\lambda=\lambda_c$
  the location of the maximum in $P(x)$ changes discontinuously.  For
  $\lambda\to 1$, $P(x)$ approaches a Gaussian.}
\label{fig:schematic}
\end{center}
\end{figure}

We numerically integrate Eq.~\eqref{PNx} by the Van Duen-Cools algorithm to
give the single-coordinate probability distribution $P_N(x)$ whose evolution
as a function of $\lambda$ is schematically illustrated in
Fig.~\ref{fig:schematic}.  Notice that there is a value $\lambda\approx
0.5567$ for which the curvature at the origin vanishes.  However, at this
value of $\lambda$ the global maximum of the $P(x)$ is not at the origin.
Thus points where $P''(x)=0$ do not help locate the global extrema of the
probability distribution and we must resort to the numerical integration.

\begin{figure}[ht]
\begin{center}
\includegraphics[width=0.7\textwidth]{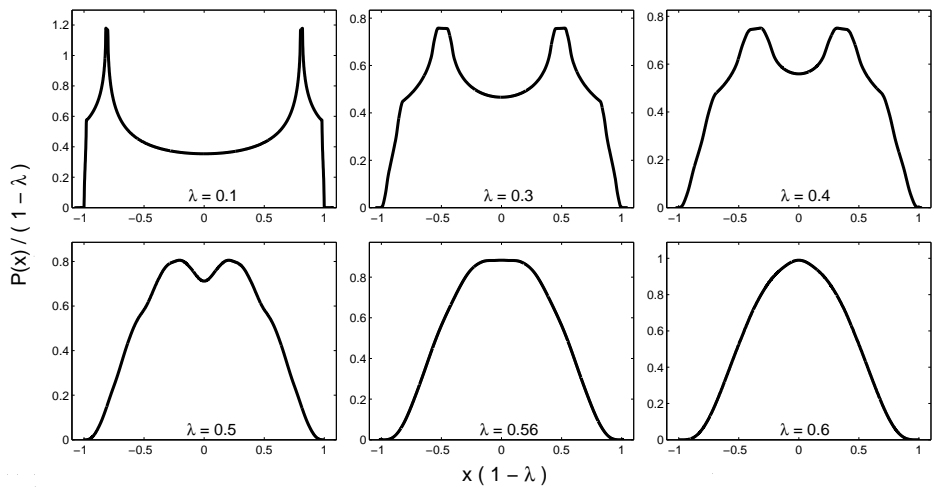}
\caption{ Scaled $x$-coordinate distribution for the shrinking Pearson walk
  in two dimensions for $\lambda=0.1$, 0.3, 0.4, 0.5, 0.56, and 0.6 (upper
  left to lower right).}
\label{fig:x}
\end{center}
\end{figure}

Since the individual step lengths decay exponentially with $N$, the
finite-$N$ distribution $P_N(x)$ quickly converges to its asymptotic
$N\to\infty$ form.  For example, for $\lambda=0.56$ (close to
$\lambda_c(x)$), the displacement of the walk after 15 steps is within
$10^{-5}$ of its final endpoint.  Hence the probability distribution is
visually indistinguishable from the asymptotic distribution on the scale of
the plots in Fig.~\ref{fig:x}.  We always use values of $N$ for each
$\lambda$ to ensure that $x_N$ is within $10^{-5}$ of its final displacement.
For small $\lambda$, $P(x)$ resembles the smile distribution of the
single-step distribution in Eq.~\eqref{qN}.  As $\lambda$ approaches
$\lambda_c$ from below, the minimum at the origin gradually fills in and
disappears for $\lambda\approx 0.56$.  For $\lambda>\lambda_c$, the
distribution develops a maximum at the origin that becomes increasingly
Gaussian in appearance as $\lambda\to 1$.

\begin{figure}[ht]
\begin{center}
\includegraphics[width=0.9\textwidth]{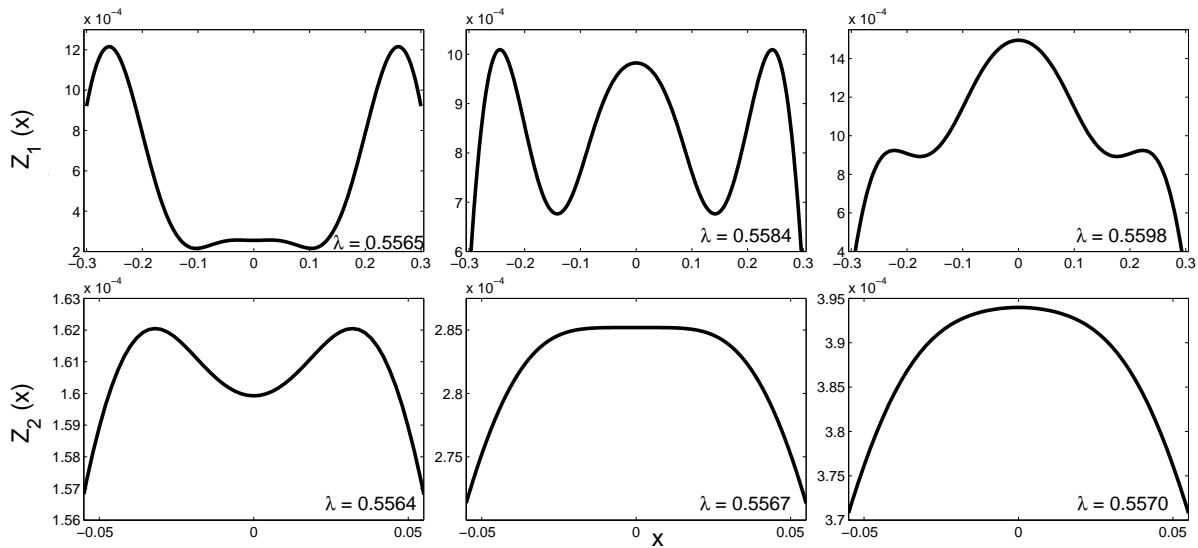}
\caption{ The single-coordinate distribution at highly-magnified scales.  Top
  line: $Z_1(x)\equiv P(x)-0.387$ for $\lambda=0.5565$, 0.5584, and 0.5598.
  Bottom line: $Z_2(x)\equiv P(x)-0.3870562$ for $\lambda=0.5564$,
  0.5567, and 0.5570. }
\label{fig:x1}
\end{center}
\end{figure}

Unexpectedly, $P(x)$ has multiple tiny maxima near the origin, that are not
visible on the scale of Fig.~\ref{fig:x}, as $\lambda$ passes through
$\lambda_c$.  The Van Duen-Cools algorithm is essential to obtain sufficient
numerical accuracy to observe these anomalies.  The top line of
Fig.~\ref{fig:x1} shows the quantity $Z_1(x)\equiv P(x)-0.387$, with the
vertical scale magnified by $10^3$ to expose the minute variations of $P(x)$.
At this magnification, one can see the birth of a maximum in $P(x)$ at the
origin that gradually overtakes the secondary maxima near $|x|\approx 0.2$.
Consequently, the location of the global maximum of $P(x)$ jumps
discontinuously from a non-zero value to zero at $\lambda =
\lambda_c(x)\approx 0.5584558\pm 0.0000003$ (as illustrated by the middle
panel on the top line of Fig.~\ref{fig:x1}, which shows $P(x)$ for
$\lambda_c-\lambda \sim \mathcal{O}\left(10^{-5}\right)$).

At a still higher resolution, the nearly flat distribution near $x=0$ at
magnification $10^{3}$ is actually oscillatory at magnification $10^{5}$
(Fig.~\ref{fig:x1} lower line).  We see that the small maximum that is born
when $\lambda$ passes through approximately 0.5565 (Fig.~\ref{fig:x1}, upper
left) actually contains an even smaller dimple that disappears when $\lambda
\gtrsim 0.5567$ (middle panel in the lower line of Fig.~\ref{fig:x1}).  To
highlight this fine-scale anomaly, we plot, in the lower line of
Fig.~\ref{fig:x1}, the quantity $Z_2(x)\equiv P(x)-0.3870562$ for three
$\lambda$ values that are very close to $\lambda_c$.  Intriguingly, we do not
find evidence of additional anomalous features at a still finer scale of
resolution.

\begin{figure}[ht]
\begin{center}
\includegraphics[width=0.8\textwidth]{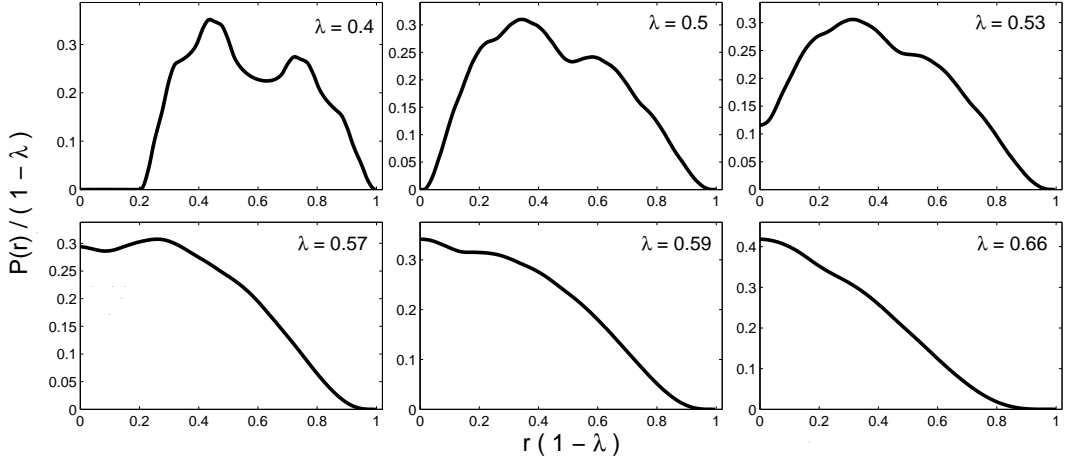}
\caption{The radial distribution for the shrinking Pearson walk in two
  dimensions for the cases $\lambda=0.40$, 0.50, 0.53, 0.57, 0.59, and
  0.66. }
\label{fig:radial}
\end{center}
\end{figure}

We also use the Van Duen-Cools algorithm to numerically integrate
Eq.~\eqref{eq.PNr} and determine the radial distribution $P_{N}(r)$.  For a
small number of steps $N$, $P_N(r)$ changes significantly with each
additional step, as was illustrated in Fig.~\ref{fig:r}.  Once the number of
steps becomes of the order of 10, however, $P_{N}(r)$ is very close to the
asymptotic $P(r)$ for $\lambda\approx \lambda_c(r)$.  The transition behavior
in $P(r)$ turns out to be much simpler than that for $P(x)$.  For $P(r)$, a
peak gradually develops at the origin, while the peak $r>0$ gradually recedes
as $\lambda$ increases.  Thus as $\lambda$ passes through $\lambda_c(r)$, the
location of the global peak of $P(r)$ discontinuously jumps to zero
(Fig.~\ref{fig:radial}).  We do not find evidence of fine-scale anomalies in
the radial distribution as $\lambda$ passes through $\lambda_c(r)$.

\section{HIGHER DIMENSIONS}
\label{sec.gend}

The approach developed for two dimensions can be straightforwardly extended
to higher spatial dimensions.  For the radial distribution in $d$ dimensions,
the single-step distribution $\mathcal{Q}_N(r)$ is now
\begin{equation}
\mathcal{Q}_N(r)=\frac{1}{r^{d-1}\Omega_{d}}\:\delta\left(r-\lambda^N\right),
\label{eq.FGS}
\end{equation} 
where $\Omega_d=2\pi^{d/2}/\:\Gamma(d/2)$ is the surface area of the unit
hypersphere in $d$ dimensions and $r=|\rv|$ is the radial distance.  The
corresponding Fourier transform is~\cite{AS2}
\begin{eqnarray}
\label{QN}
  \mathcal{Q}_N(k) &=& \Omega_d^{-1} \int\dr\ \frac{\delta\left(r-\lambda^n\right)}{r^{d-1}}
  \:e^{i\kv\cdot\rv},\nonumber\\
  &=&\frac{\Gamma\left(\frac{d}{2}\right)}{\Gamma\left(\frac{1}{2}\right)
    \:\Gamma\left(\frac{d-1}{2}\right)}\int_{0}^{\pi}\text{d}\theta\:\sin^{d-2}
  \theta\:e^{ik\lambda^N\cos\theta}
  ={}_0{F}{_1}(d/2,-k^2\lambda^{2N}/4),\label{eq.Fpp}
\end{eqnarray}
where $_0F_1(a,z)$ is the confluent hypergeometric function.  The Fourier
transform $P_N(k)$ is then the product of Fourier transform of the
single-step distributions, and its Fourier inverse gives $P_N(r)$.  By
integrating over the $d-2$ azimuthal angles, and then integrating over the
polar angle $\theta$, as in Eq.~\eqref{QN}, the formal solution is
\begin{eqnarray}
P_N(r) &=& \frac{\Omega_{d-1}}{(2\pi)^d}\int dk\:
k^{d-1}P_N(k)\int_{0}^{\pi}d\theta\:\sin^{d-2}\theta\:
e^{-ikr\cos\theta},\nonumber \\
 &=& \frac{2^{1-d}}{\pi^{d/2}\Gamma(d/2)}\int_{0}^{\infty}dk\:k^{d-1}{}_0{F}{_1}(d/2,-k^2r^2/4)\: \:\prod_{n=0}^{N}{}_0{F}{_1}(d/2,-k^2\lambda^{2n}/4)\,.
\end{eqnarray}

Since $_0F_1(\nu+1,-(z/2)^2) \propto (z/2)^{-\nu}J_\nu(z)$~\cite{AS2}, we can
again numerically determine $P_N(r)$ by using the Van Duen-Cools algorithm.
The result of this calculation is that the radial distribution undergoes a
second-order transition at $\lambda_c$ in which the location of the single
maximum continuously decreases to zero as $\lambda$ increases beyond
$\lambda_c$.

The same formal approach can be used to calculate the distribution $P(x)$.
This distribution now remains peaked at the origin for all values of
$\lambda$.  The physical origin of this property stems from the nature of the
single-step distribution.  The generalization of Eq.~\eqref{qN} is 
\begin{equation*}
q_n(x) \propto \left[\lambda^{2(n-1)}-x^2\right]^{(d-3)/2}~.
\end{equation*}
This function is flat for $d=3$ and peaked at the origin for $d>3$.
Consequently, the convolution of these single-step distributions leads to
$P_N(x)$ having a single peak at the origin.

\section{DISCUSSION}
\label{sec.conv}

We investigated the shrinking Pearson walk, where each step is in a random
direction, while the length of the $n^{\rm th}$ step is $\lambda^{n-1}$, with
$\lambda<1$.  Because the step lengths are not identical, one of the defining
conditions for the central limit theorem is violated.  Consequently, there is
no reason to expect that the probability distribution for this walk is
Gaussian.  We studied basic properties of the radial probability
distribution, $P(r)$, and the distribution of a single coordinate, $P(x)$.
Because a walk with a large displacement in one direction necessarily implies
a small displacement in the orthogonal direction, $P(r)$ does not simply
factorize as a product of single-coordinate distributions.  The $P(r)$ and
$P(x)$ are distinct distributions.

In two dimensions, the radial probability distribution of the shrinking
Pearson walk changes from being peaked away from the origin to being peaked
at the origin as the shrinking factor $\lambda$ increases beyond a critical
value $\lambda_c(r)$.  As this transition in $\lambda$ is passed, the
location of the peak changes discontinuously from a non-zero value to $r=0$.
In greater than two dimensions, there is a similar shape transition in the
radial distribution, but now the location of the only peak goes to zero
continuously as $\lambda$ increases beyond $\lambda_c(r)$.

The single-coordinate distribution $P(x)$ has peculiar features for the
specific case of two dimensions.  Visually, $P(x)$ becomes nearly flat at the
origin for $\lambda\approx 0.5565$ (middle panel, bottom row of
Fig.~\ref{fig:x}).  However, at a higher degree of magnification, this nearly
flat portion of the distribution exhibits fine-scale oscillations, with up to
seven local extrema.  Because additional oscillations can be resolved as the
resolution is increased, it is tempting to speculate that arbitrarily many
oscillations occur at progressively decreasing scales.  To test for this
possibility, we computed the first derivative $P_N'(x)$ from Eq.~\eqref{PNx},
and looked for additional zeros in $P_N'(x)$ as a function of $x$.  Again
employing the Van Duen-Cools algorithm, we find the $P_N(x)$ is strictly
positive for $x$ in the range $5\times 10^{-8}$ to $10^{-4}$ when
$\lambda=0.55672$, but is strictly negative in the same range of $x$ when
$\lambda=0.55673$.  Moreover, $P_N'(x)$ appears to scale as $x^{1/2}$ in the
range $5\times 10^{-8}< x< 10^{-4}$, so we anticipate no additional zeros for
$x\to 0$.  This numerical test suggests that there are no additional
oscillations in $P(x)$ beyond those revealed in Fig.~\ref{fig:x1}.

\medskip {We are grateful for financial support from DOE grant
 DE-FG02-95ER14498 (CAS) and NSF grant DMR0535503 and DMR0906504 (SR).}


\begin{thebibliography}{99}
  
\bibitem{pearson} K. Pearson, Nature {\bf 72} 294; 318; 342 (1905).

\bibitem{W94} G. H. Weiss, {\it Aspects and Applications of the Random Walk}
  (North-Holland, Amsterdam, 1994).

\bibitem{W} B. Jessen and A. Wintner, 
  Trans.\ Amer.\ Math.\ Soc.\ {\bf 38}, 48--88 (1935); B. Kershner and
  A. Wintner,
  Amer.\ J. Math.\ {\bf 57}, 541--548 (1935); A.  Wintner,
  {\it ibid.} {\bf 57}, 827--838 (1935).

\bibitem{E} P. Erd\H os,
  Amer.\ J. Math.\ {\bf 61}, 974--976 (1939); P. Erd\H os, 
{\it ibid.} {\bf 62}, 180--186 (1940).

\bibitem{G} A. M. Garsia, 
  Trans.\ Amer.\ Math.\ Soc.\ {\bf 102}, 409--432 (1962); A. M. Garsia,
  Pacific J. Math.\ {\bf 13}, 1159--1169 (1963).
  
\bibitem{Kac} M. Kac, {\it Statistical Independence in Probability, Analysis
    and Number Theory} (Mathematical Association of America; distributed by
  Wiley, New York, 1959).
    
\bibitem{KR} P.~L.~Krapivsky and S.~Redner, Am.\ J. Phys.\ {\bf 72}, 591--598
  (2004).

\bibitem{PSS} Y. Peres, W. Schlag, and B. Solomyak,
  in {\it Fractals and Stochastics II}, edited by C. Bandt, S. Graf, and
  M. Z\"ahle (Progress in Probability, Birkhauser, 2000), vol.\ 46, pp.\
  39--65.

\bibitem{BS99} E. Barkai and R. Silbey, Chem.\ Phys.\ Lett.\ {\bf 310}, 287
  (1999); Phys.\ Chem.\ B, {\bf 104}, 342 (2000).

\bibitem{WK83} G. H. Weiss and J. E Kiefer, J. Phys.\ A {\bf 16}, 489--495
  (1983).

\bibitem{R06} T. Rador, Phys.\ Rev.\ E {\bf 74} 051105 (2006).

\bibitem{MK07} S.~N. Majumdar and M.~J. Kearney, Phys.\ Rev.\ E, {\bf 76},
  031130 (2007).
  
\bibitem{bazant} M. Bazant, private communications.  See, also lecture notes
  by M. Bazant for MIT course 18.366. The URL is {\tt
    {<}http://ocw.mit.edu/Ocw/Mathematics/18-366Fall-2006/CourseHome/{>}}.

\bibitem{vdc} J. Van Deun and R. Cools, 
  ACM Transactions on Mathematical Software {\bf 32} 580--596 (2006); Comp.\
  Phys.\ Comm.\ {\bf 178} 578--590 (2008).

\bibitem{NL} E. Ben-Naim, S. Redner, and D. ben-Avraham, Phys.\ Rev.\ A {\bf
    45}, 7207 (1992); D. ben-Avraham, F. Leyvraz, and S. Redner, Phys.\ Rev.\
  A {\bf 45}, 2315 (1992).

\bibitem{AS} M. Abramowitz and I. A. Stegun {\it Handbook of Mathematical
    Functions} (Dover, New York, 1972).  See 9.1.21.

\bibitem{AS1} See 10.1.1. anmd 10.1.11 in reference~\cite{AS} that gives the
  representation of $\cos z$ in terms of Bessel functions.

\bibitem{AS2} See 9.1.20 and 9.1.69 in reference~\cite{AS} for the connection
  between the relevant Fourier integrals and the hypergeometric function.

\bibitem{oss} See for example: L. N. G. Filon, Proc.\ Roy.\ Soc.\ Edinburgh
  {\bf 49} 38-49 (1928), Y. L. Luke, Proc.\ Cambridge Phil.\ Soc.\ {\bf 50},
  269--277 (1954), B. Gabutti, Math.\ Comp.\ {\bf 33}, 1049-1057 (1979),
  G. A. Evans, {\it Practical Numerical Integration}, (Chaps.\ 3 and 4)
  (Wiley, New York, 1993).

\bibitem{b3} S. Groote, J. G. K\"{o}rner, and A. A. Pivovarov, 
Nucl.\ Phys.\ B {\bf 542}, 515--547 (1999).

\bibitem{b1} S. Davis, 
Class.\ Quantum Grav.\ {\bf 18}, 3395--3425 (2001).

\bibitem{b2} R. Gaspard and D. Alonso Ramirez, 
Phys.\ Rev.\ A {\bf 45}, 8383--8397 (1992).

\bibitem{b4} T. Lotter, C. Benien, and P. Vary, 
  EURASIP Journal on Applied Signal Processing {\bf 11}, 1147--1156 (2003).

\end{thebibliography}
\end{document}